\documentclass[aps,prd,showpacs,nofootinbib]{revtex4}
\usepackage{latexsym}
\usepackage{amsfonts}
\usepackage{amsbsy}
\usepackage{mathrsfs}
\usepackage{graphicx,calc,epsfig}
\def\ut#1{\rlap{\lower1ex\hbox{$\sim$}}#1{}}
\newcommand{\N}{\mathbb{N}}

\newcommand{\R}{\mathbb{R}}
\newcommand{\be}{\nopagebreak[3]\begin{equation}}
\newcommand{\ee}{\end{equation}}
\newcommand{\ba}{\nopagebreak[3]\begin{eqnarray}}
\newcommand{\ea}{\end{eqnarray}}
\DeclareFontFamily{U}{rsfs}{}         
\DeclareFontShape{U}{rsfs}{m}{n}{<5> rsfs5 <6><7> rsfs7          %
  <8><9><10><10.95><12><14.4><17.28><20.74><24.88> rsfs10}{}     %
\DeclareMathAlphabet{\mathfs}{U}{rsfs}{m}{n}                     %
\newcommand{\mfs}[1]{\mathfs {#1}}                               %
\newcommand{\va}{\scriptscriptstyle}

\newcommand{\sF}{{\mfs F}}

\newcommand{\sH}{{\mfs H}}

\newcommand{\sS}{{\mfs S}}
\newcommand{\sM}{{\mfs M}}

\newcommand{\sW}{{\mfs W}}

\newcommand{\g}{\mathfrak{g}}

\newcommand{\tr}{\mathrm{tr}}

\begin{document}

\title{Two-dimensional topological field theories coupled to four-dimensional BF theory}

\date{\today}

\author{Merced Montesinos\footnote{Associate Member of the Abdus Salam
International Centre for Theoretical Physics, Trieste, Italy.}}

\email{merced@fis.cinvestav.mx} \affiliation{Departamento de F\'{\i}sica,
Cinvestav, Avenida Instituto Polit\'ecnico Nacional 2508, San Pedro Zacatenco,
07360, Gustavo A. Madero, Ciudad de M\'exico, M\'exico}

\author{Alejandro Perez}
\affiliation{Centre de Physique Th\'eorique\footnote{Unit\'e Mixte
de Recherche (UMR 6207) du CNRS et des Universit\'es Aix-Marseille
I, Aix-Marseille II, et du Sud Toulon-Var; laboratoire afili\'e
\`a la FRUMAM (FR 2291)}, Campus de Luminy, 13288 Marseille,
France.\\ and\\
Universidade Federal do Espiritu Santo, Campus Goiabeiras,
29060-900 Vitoria, Brasil.}

\begin{abstract}
Four dimensional BF theory admits a natural coupling to extended
sources supported on two dimensional surfaces or string
world-sheets. Solutions of the theory are in one to one
correspondence with solutions of Einstein equations with
distributional matter (cosmic strings). We study new (topological
field) theories that can be constructed by adding extra degrees of
freedom to the two dimensional world-sheet. We show how two
dimensional Yang-Mills degrees of freedom can be added on the
world-sheet, producing in this way, an interactive (topological)
theory of Yang-Mills fields with BF fields in four dimensions. We
also show how a world-sheet tetrad can be naturally added. As in
the previous case the set of solutions of these theories are
contained in the set of solutions of Einstein's equations if one
allows distributional matter supported on two dimensional
surfaces. These theories are argued to be exactly quantizable. In
the context of quantum gravity, one important motivation to study
these models is to explore the possibility of constructing a
background independent quantum field theory where local degrees of
freedom at low energies arise from global topological
(world-sheet) degrees of freedom at the fundamental level.
\end{abstract}


\maketitle

\section{Introduction}

Topological field theories are simple examples of background
independent field theories for which quantization can be
completely worked out. These theories are a natural play ground
where conceptual as well as technical issues in background
independent quantum theory can be addressed in detail.
Three-dimensional vacuum general relativity is an important
example of a topological field theory. Interestingly the
topological nature of the theory can be maintained if matter is
added in the form of topological defects representing massive and
spinning point particles \cite{1}. Interest in the quantization of
$2+1$ gravity coupled to point particles has been revived in the
context of the spin foam \cite{spinfoam} and loop quantum gravity
\cite{lqg} approaches to the nonperturbative and
background-independent quantization of gravity.  On the one hand
this simple system provides a nontrivial example where the strict
equivalence between the covariant and canonical approaches can be
demonstrated \cite{ky}. On the other hand intriguing relationships
with field theories with infinitely many degrees of freedom have
been obtained \cite{lau,Noui:2006kv}. The generalization of these
models to higher dimensions has been studied in \cite{bp}. As it
is shown there, membrane-like defects of dimension $d-3$ are a
natural form of matter that couples to $d$-dimensional BF theory
\cite{Horowitz:1989ng}. The resulting theory is in turn also a
topological theory and can be completely quantized using the
techniques of loop quantum gravity. Among these higher dimensional
models the four-dimensional one (which couples to string-like
defects) is of singular interest due to the special role played by
$4$-dimensional BF theory in the construction of spin foam models
of four-dimensional quantum gravity.

At first look these strings are a rather dull form of matter: at
their location there are conical singularities of the curvature
tensor and the equations of motion imply that the string world
sheet is locally flat \cite{winston} (vibrational modes of the
strings are pure gauge). Nevertheless, as we will argue in this
paper, the feature that makes these strings interesting is the
fact that they are extended objects (this is also behind their
exotic statistical properties \cite{baez}). This will allow us to
couple four-dimensional BF theory with more physically appealing
degrees of freedom. As the set of possibilities is quite vast, we
will restrict our attention to certain world-sheet theories that
satisfy the following two properties: (a) they can be naturally
(or minimally) coupled to BF theory in 4d, and (b) the coupled
system defines a (topological) theory with no local degrees of
freedom. Due to the close relationship between four-dimensional BF
theory and gravity, requirement (a) is expected to produce
physically interesting models, as they might provide natural
candidates for the coupling of spin foam models of gravity with
natural forms of matter. Requirement (b) implies that the models
studied here are expected to be non-perturbatively quantizable.

 We believe that the study of these simple topological models can
be of more relevance than a simple exercise in the application of
non-perturbative quantization techniques.  We would like to
explore the possibility that topological theories, containing low
dimensional objects, could be used to construct a background
independent quantum field theory with infinitely many
(`quasi-local') degrees of freedom. This is in fact our
motivation for imposing requirement (b) above.

The article is organized as follows: In Section \ref{strings} we
briefly review the coupling of strings to four-dimensional BF
theory. In Section \ref{ymbf} we show how Yang-Mills degrees of
freedom can be added to the strings. We analyze the equations of
motion of the coupled system and perform the canonical analysis to
prove that the theory is topological. In Section \ref{gr} we add a
tetrad field on the world sheet and obtain an interesting model
whose equations of motion resemble those of general relativity in
a curious way. In Section \ref{2dym} we study a purely
two-dimensional model of background independent Yang-Mills theory
which naturally follows from the results of the previous sections.
In Section \ref{especulemos} we present a speculative discussion
about the possibility of using topological theories of the type
introduced in this article in order to define a background
independent quantum field theory with infinitely many degrees of
freedom..

\section{Strings coupled to four-dimensional BF theory}\label{strings}

The coupling of $(d-3)$-dimensional membranes to $d$-dimensional
BF theory (defined for a large class of structure groups) was
introduced in \cite{bp}. Here we concentrate on the case of
strings coupled to four-dimensional BF theory with structure group
$SO(3,1)$ (see Refs. \cite{mm} for its canonical analysis and
Refs. \cite{mon,mon2} for alternative action principles). If we
denote $\sM$ the four-dimensional space time manifold and
$\sW\subset \sM$ the two-dimensional world sheet of the string,
the action defining the coupling is given by \be \label{membrane}
S_{\va ST-BF}= \int_{\sM} B_{IJ} \wedge F^{IJ}(A) + {\tau}
\int_{\sW} (B + d_Aq)^{IJ}p_{IJ}, \ee where $I,J=1,..,4$, and if
we denote $T_{IJ}\in so(3,1)$ the generators of the Lie algebra
then
 $q=q^{IJ}T_{IJ}$ is a $so(3,1)$-valued
$1$-form on $\sW$ and $p=p^{IJ}T_{IJ}$ is a $so(3,1)$-valued function on $\sW$.
This action is invariant under the gauge transformations: \be
\label{gravy3}
\begin{array}{ccl}
&& B \mapsto  gBg^{-1} \ \ \ \ \ \ \ \ \ \ \ \ \ \ \ \ \ \ \ \ \ \ \  B\mapsto B+d_A\eta \\
&& A \mapsto gAg^{-1}+g dg^{-1} \ \ \ \ \ \ \ \ \ \ \ \ q \mapsto q -\eta \\
&& q \mapsto  gq g^{-1} \\
&& p \mapsto gpg^{-1},
\end{array}
\ee where $g\in  C^{\infty}(\sM,G)$ and $\eta$ is any $\g$-valued
$(d-3)$-form. Varying the action with respect to the $B$ field
implies that the connection $A$ is flat except at $\sW$: \be F =
-p\ \delta_{\sW} , \ee where $\delta_{\sW}$ is the distributional
2-form (current) associated to the string world-sheet. So, the
string causes a conical singularity in the otherwise flat
connection $A$. The strength of this singularity is determined by
the field $p$, which plays the role of a `momentum density' for
the string.  Note that while the connection $A$ is singular in the
directions transverse to $\sW$, it is smooth and indeed flat when
restricted to $\sW$. Thus the equation of motion obtained from
varying $q$ makes sense: \be d_A p = 0. \ee This expresses
conservation of momentum density and in fact implies that the
field $p$ remains in the same conjugacy class, hence it can be
writen as $p=\tau \lambda v\lambda^{-1}$ for $v\in so(3,1)$ a
normalized vector and $\lambda\in SO(3,1)$. The constant $\tau$
defines the string tension. Conjugacy classes of $so(3,1)$ are
labelled by the two Lorentz Casimirs. So far we have fixed only
one by choosing the string tension $\tau^2=p_{IJ}p^{IJ}$. The
other Casimir  defines an extra parameter
$s=p_{IJ}p_{KL}\epsilon^{IJKL}$ (the geometric meaning of $s$ will
be discussed below). Notice that the strength of the conical
singularity at the location of the strings is in this sense non
dynamical. This will change in the model of Section \ref{ymbf}.

 Assuming the spacetime manifold is of
the form ${\sM}=\Sigma \times \R$.   We choose local coordinates
$(t,x^a)$ for which $\Sigma$ is given as the hypersurface
$\{t=0\}$.  By definition, $x^a$ with $a=1,2,3$ are local
coordinates on $\Sigma$. We also choose local coordinates $(t,s)$
on the $2$-dimensional world-sheet $\sW$, where $s \in [0,2\pi]$
is a coordinate along the one-dimensional string formed by the
intersection of $\sW$ with $\Sigma$. Performing the standard
Legendre transformation one obtains $E^a_i=\epsilon^{abc} B_{ibc}$
as the momentum canonically conjugate to $A_{b}^i$. Similarly,
$p_{IJ}$ is the momentum canonically conjugate to
$q^{IJ}_{1}=q^{IJ}_a(\partial_{\sigma})^{a}$. The phase space
variables satisfy the following constraints: \be \label{gauss}
L_{IJ}:=D_{a}E^{a}_{IJ} - 2 \delta_{{\sS}}[ q_{1[I|M|} p^M_{\ J]}]
\approx 0 \ee \be \label{curvature} K^{a}_{IK}:=\epsilon^{a b
c}F^{IJ}_{bc}(x) + \delta_{{\sS}}[p^{IJ}
(\partial_{\sigma})^{a}]\approx 0, \ee Here $\sS\subset\Sigma$
denotes the one-dimensional curve representing the string,
parametrized by $x_{\va \sS}(s)$, and for any field $\phi$ on
$\sS$ we define
\[\delta_{{\sS}}[\phi]:=\int_{\sS} \phi\ \delta^{\va (3)}(x- x_{\va \sS}(s)).\]
The constraint (\ref{gauss}) is the
modified Gauss law of $BF$ theory due to the presence of the
string. The constraint (\ref{curvature}) is the modified curvature constraint
containing the dynamical information of the theory.  This
constraint implies that the connection $A$ is flat away from the
string $\sS$.    Some care must be taken to correctly intepret the
constraint for points on $\sS$.  By analogy with the case of 3d
gravity, the correct interpretation is that the holonomy of an
infinitesimal loop circling the string at some point $x \in \sS$
is $\exp(-p(x)) \in G$, where $p = \tau \lambda v \lambda^{-1}$ as
before.  This describes the conical singularity of the connection
at the string world-sheet.

The $BF$ phase space variables satisfy the standard commutation
relations: \be \{E_i^{a}(x),A_{b}^j(y)\}=\delta_{b}^{a}\delta_i^j
\, \delta^{\va(3)}(x-y) \ \ \ \ \
\{E_i^{a}(x),E^{b}_j(y)\}=\{A_{a}^i(x),A_{b}^j(y)\}=0. \ee The
phase space of the string is parametrized in terms of the momentum
$p^{IJ}$ and the `total angular momentum' $J_{IJ} =2 q_{1[I|M|}
p^M_{\ J]}$. The Poisson brackets of these variables are given by
\be \{p_{IJ}(s),J_{KL}(s')\} = c_{IJKL}^{ST} p_{ST}(s) \delta^{\va
(1)}(s-s')\ \ \ \ \ \{J_{IJ}(s),J_{KL}(s')\} = c_{IJKL}^{ST}
J_{ST}(s) \delta^{\va (1)}(s-s'), \ee where $c_{IJKL}^{ST}$ are
the structure constants of $so(3,1)$, and \be \label{***}
\{J_{IJ}(s),\lambda(s')\} = -T_{IJ}\lambda(s) \delta^{\va
(1)}(s-s'). \ee The string variables are still subject to the
following first class constraints: \be \label{spin}
\tr[T_{IJ}\lambda z\lambda^{-1}]J^{IJ}=0 \ \ \ \ \ \tr[p \lambda
z\lambda^{-1}] = \tau \tr[v z], \ee where $z\in\g$ is such that
$[z,v]=0$. The last constraint is the
generalization of the mass shell condition for point particles in
3d gravity.
The Poisson bracket of the string variables with the $BF$
variables is trivial, as well as the Poisson brackets among the
$p_{IJ}$.

\subsection{Geometrical interpretation}\label{geom}

Here we  present a brief account of the analysis  carried out in
\cite{winston}. The full set of equations of motion of the theory
is
\be \nonumber \begin{array}{ccc}F(A)=-p\ \delta_{\sW} \\
d_A B= -[q, p] \ \delta_{\sW}\label{imp} \\
d_A p|_{\sW}=0, \  \phi^*_{\va \sW}(B+d_Aq)=0\end{array}, \ee
where $\phi^*_{\va \sW}$ in the last equation denotes the pull
back of the corresponding 2-forms to $\sW$.  Therefore, the field
configurations $A=0$, $B=0$, $q=0$, $p=${\em constant} gives a
solution to the equations of motion in an open region $U\subset M$
such that any open set containing points of $\sW$ has points
outside $U$. Since the theory is topological, all the solutions
are equivalent to this one in $U$ through a gauge transformation.
Assume that we have a coordinate system in $U$ with coordinate
functions $X^I$, (for $I=1,\cdots, 4$). In order to recover an
interpretation of fields on a flat background we can make a gauge
transformation of the type (\ref{gravy3}) with gauge parameter
$\eta^{IJ}=X^{[I}dX^{J]}$. In this gauge the solution is \be
B_{ab}^{IJ}=e^I_{[a}\ e_{b]}^J=\delta^I_{[a}\delta^J_{b]} \ \
q_a^{IJ}=X^{[I}d_aX^{J]}.\ee We see that in this gauge the $B$
field defines a flat background geometry. There is still the
residual gauge freedom that maintains this property of the $B$
field given by gauge transformations of the form $\eta_0=df$ for
some arbitrary $f$. We call this family of gauges {\em flat
gauges}. The integrability conditions that follow from the
equation $dB=[q,p] \delta_{\sW}$ imply that $d[p,q]=0$ or
equivalently that $[p,q]=d\alpha$ for some potential $\alpha$. If
$\alpha=0$ it can be shown that $[p,q]=0$ has non trivial
solutions if $s=p_{IJ}p_{KL}\epsilon^{IJKL}=0$ \footnote{If we
allow for complex $p^{IJ}$, then solutions exist also if $p^{IJ}$
is self-dual (or anti self-dual).}. In that case the string
world-sheet $X^I(\sigma,t)$ is given by a plane in Minkowski
spacetime passing through the origin defined by either the
equation $p^{IJ}X_I=0$ or $\star p^{IJ}X_I=0$. We can translate
the plane off the origin by choosing $\alpha^{IJ}=C^{[I}X_{}^{J]}$
(this choice sends $X^I$ to $X^I+C^I$). If $s\not=0$ then equation
$[p,q]=0$ implies $X^I=0$.

 One can establish a strict
connection between these solutions and solutions of general
relativity representing a cosmic string. In cylindrical
coordinates $\{\partial_t,
\partial_r,
\partial_{\varphi}, \partial_z\}$ such that the string is lying along the $z$ axis and
goes through the origin the metric of a cosmic string solution of
tension $\tau$ is: \be \label{cosmic}
ds^2 = g_{\mu \nu} dx^{\mu} \otimes dx^{\nu} \\
     = -dt^2 + dr^2 + (1-a)^2 r^2 d\varphi^2 + dz^2 ,
\ee where $a = (1-4G\tau)$, $G$ is the Newton constant. The dual
co-frame for the above metric is written \be e^0 = dt \ \ \ \ e^1
= \cos \varphi d r - a\ r \sin \varphi d \varphi \ \ \ \ e^2 =
\sin \varphi d r + a\ r \cos \varphi d \varphi \ \ \ \ e^3 = dz,
\ee such that $ds^2=e^I \otimes e^J \eta_{IJ}$. The spin
connection  (s.t. $d_A e=0$) is \be A = A^{IJ}_{\mu} \, J_{IJ} d
x^{\mu} = 4 G \tau \, J_{12} \, d \varphi, \ee where $J_{IJ}$ are
the  $so(3,1)$ generators. We can identify now the string momentum
$p$ above, namely $p^{IJ}J_{IJ}=\tau J_{12}$. From the
distributional identity $d d \varphi = 2 \pi \delta^2(r) dx dy$
($x=r \cos \varphi$, $y=r \sin \varphi$), it is immediate to
compute the torsion $T = T^0 e_0$ and curvature $F = F^{12} \,
\sigma_{12}$ of the cosmic string induced metric: \be T^0 = 0,\ \
\ F^{12} = 8 \pi G \tau \, \delta^2(r) \, dx dy . \ee The above
fields are clearly a solution of Einstein's equations  with
distributional matter \be \epsilon_{IJKL} e^J \wedge F^{KL} = 8
\pi G \tau \, \epsilon_{IJKL} e^J J_{12}^{IJ} \, \delta_{\sW}. \ee
The previous solution is in one to one correspondence with the
solution of (\ref{imp}) \be B=(e\wedge e)^*, \ \ \  A =  4 G \tau
\, J_{12} \ d\varphi, \ \ \  p= \tau \, J_{12},\ \ \
q^{IJ}=\left(z dt -t dz\right)\delta^{[I}_0\delta^{J]}_{3}=\left(z
dt -t dz\right)J_{21}^{IJ}.\ee One can construct a two string
solution by `superimposing' two solutions of the previous kind at
different locations (notice that the equations are non linear so
the new solution is not the sum of two solutions). It can be show
that the torsion $d_AB$ is proportional to the distance separating
the worldsheets in the flat-gauge where
$B^{IJ}_{ab}=\delta^{[I}_a\delta^{J]}_b$. More strings can be
added in a similar fashion.

\section{Minimal coupling of world sheet Yang-Mills with 4d BF
theory}\label{ymbf}

Yang Mills theory in two dimensions can be written in a way that
resembles BF theory if one is given a $2$-form field $\rho$,
namely
\begin{eqnarray}\label{ym}
S_{\va YM} =\int_{\sW} \left [ {\cal E}_{a} F^{a}(A)+\rho {\cal
E}_{a}{\cal E}^{a} \right ],
\end{eqnarray}
where  $a=1,\ldots, {\rm dim}({\frak g})$ are internal indices
labeling the elements of a basis of the Lie algebra $\frak g$ of
the gauge group of our choice $G$ (we require $G$ to be compact
and $\frak g$ to have an invariant metric with which we raise and
lower internal indices). The field $A= \left ( A^{a}_{\mu} d
x^{\mu} \right ) \otimes J_a$ is the $\frak g$-valued connection
1-form, $[J_a , J_b]= f^c\,_{ab} J_c$ where $f^c\,_{ab}$ are the
structure constants with respect to the basis $\{J_a \}$. AUnder
these assumptions the internal metric can  be taken as $k_{ab}=c
\mbox{Tr} \, J_a J_b$ (assuming a matrix form for the generators
$J_a$ and $c$ is a constant that depends on the dimension of the
representation of the $J_a$). The field ${\cal E}_{a}$ is a
collection of ${\rm dim}(\frak g)$ many $0$-forms. One can show
that if $\rho$ is non-degenerate (i.e., a volume form) the
previous action is equivalent to the standard Yang Mills action
\[S_{\va YM}=\int_{\sW} \sqrt{g} g^{\mu\nu}g^{\rho\rho}F^{a}_{\mu\nu}F_{\rho\rho\, a},\]
where the 2d metric $g_{\mu\nu}$ is such that $\rho=\sqrt{g}
dx^1\wedge dx^2$. If one does the canonical analysis of the BF
like action above one finds that the total Hamiltonian is not
weakly vanishing due to the presence of the background structure
provided by the (non-dynamical) $\rho$. It is also easy to check
through the canonical analysis that the theory has no local
degrees of freedom. Sometimes it is said that 2d YM is
topological; however, this is not strictly the case because, even
though the degrees of freedom are global (and certainly tied to
the topology of $\sW$), they are also related to the background
structure $\rho$.

The simplest way of coupling two-dimensional Yang Mills theory
with four-dimensional BF theory to produce a background
independent field theory is to combine the $B$ field and the world
sheet variable $p$ to produce a volume $2$-form
$\rho=B^{IJ}p_{IJ}$ on the world sheet. The result is given by the
following action:
\begin{eqnarray}\label{bfym}
S_{\va BFYM}&=& \int_{\sM} B_{IJ}\wedge F^{IJ}(\omega)+\int_{\sW} \left (
\left [ B^{IJ}{\cal E}_a{\cal E}^a -d_{\omega} q^{IJ} \right ]
p_{IJ} +{\cal E}_aF^a(A)\right )
\end{eqnarray}
The equations of motion of the new model are \be F(\omega)
+\delta_{\sW}[{\cal E}_a{\cal E}^ap]=0,\ \ \ \ d_{\omega} B+
\delta_{\sW}[qp]=0, \ \ \ \phi^*_{\sW}({\cal E}_a{\cal
E}^aB-d_{\omega}q)=0,\ee and \be 2B\cdot p \ {\cal E}^a+F^{a}(A)=0
.\ee We have not explicitly written the equations $d_{\omega}p=0$,
and $d_A{\cal E}^a=0$ as they implied by the integrability
conditions arising from the Bianchi identities for the curvature
of $\omega$ and $A$, respectively.

Now we show that the new model is in fact a topological field
theory (i.e. background independent with no local degrees of
freedom). In order to do this we perform the 3+1 decomposition of
the previous action and analyze its phase space structure. The
unconstrained phase space is parametrized by the canonical
variables $(E^{\mu}_{IJ}, A_{\nu}^{KL})$ and $(p^{IJ}, q_1^{KL})$
(of the previous section) plus the Yang-Mills canonical pair
$({\cal E}_a,{\cal A}_1^b)$.  The constraints relating the bulk
degrees of freedom with the ones on the world sheet are  \be
\label{gauss1} L_{IJ}:=d_{A{\mu}}E^{\mu}_{IJ} + 2 \delta_{{\sS}}[
q_{[I|M|} p^M_{\ J]}]\approx 0 \ee \be \label{curvature1} K^{\mu
IJ}:=\epsilon^{\mu\nu\rho}F^{IJ}_{\nu\rho}(x) +
\delta_{{\sS}}[{\cal E}_a{\cal E}^a p^{IJ}\partial^{\mu}_{\sigma}]
\approx 0.\ee Notice that $L_{IJ}$ is precisely the same as
(\ref{gauss}), while $K^{IJ}$ is a simple modification of
(\ref{curvature}). In fact there are new constraints
\begin{eqnarray}
{G}_a &:=& d_{\cal A}{\cal E}_a\approx 0 , \label{cont}
\end{eqnarray}
which is the standard Gauss law of Yang-Mills. These equations
(together with Hamilton's equations of motion) imply that ${\cal
E}_a{\cal E}^a=constant$. It is easy to see that the constraint
algebra closes forming a first class system of $6+18+{\rm
dim}(\frak{g})$ local constraints for the same number of
configuration variables $\{q_1^{IJ},A_{\mu}^{IJ},{\cal A}_1^a\}$.
The model has no local degrees of freedom \footnote{There is a
subtlety concerning the constraints $K_{IJ}$. In fact when we are
away from the string the source term vanishes and the Bianchi
identity implies that only $3$ out of the $6$ ones are
independent. On the string the Bianchi identity implies $d_{A}p=0$
which is indeed and independent condition.}. The curvature
constraint implies that the space-time connection is flat in the
bulk and there is a conical singularity at the string. The strings
on $\Sigma$ can be viewed as flux lines of Yang-Mills electric
field which back react with the environment producing a conical
singularity whose strength is modulated by the Yang-Mills `energy
density' $\rho_{\cal E}= \delta_{{\sS}}[{\cal E}_a{\cal E}^a
p_{IJ}]$. As mentioned in the introduction the strength of the
curvature singularity is now dynamical.

\section{Adding a world sheet `frame' fields} \label{gr}

The idea follows from the observation that the two-dimensional field theory
defined by the following action has no local degrees of freedom
\begin{eqnarray}\label{toy}
S&=& \int_{\sW} \left ( \left [ d q^{IJ} + \ast (e^I \wedge
e^J)\right ] p_{IJ} + \pi_I d e^I \right )
\end{eqnarray}
where $\ast (e^I \wedge e^J)= \frac12 \varepsilon^{IJ}\,_{KL} e^K
\wedge e^L $, $\sW$ is a two-dimensional surface, $q^{IJ}=-
q^{JI}$ is a set of six 1-forms on $\sW$, $e^I$ is a set of four
1-forms on $\sW$, $p_{IJ}=-p_{JI}$ is a set of six 0-forms
(functions) on $\sW$, $\pi_I$ is a set of four 0-forms (functions)
on $\sW$. In principle, there are other terms that can also be
added to the action, for instance, $( d \ast q^{IJ}) p_{IJ}= d
q^{IJ} \ast p_{IJ}$ and $(e^I \wedge e^J) p_{IJ}$.

In order to count the number of degrees of freedom let us perform
the canonical analysis of this model.  Let
$(y^a)=(y^0,y^1)=(\tau,\sigma)$ be local coordinates on $\sW$
which is assumed to have the form $\sW= {\sS} \times \mathbb{R}$;
the coordinate time $\tau$ labels the points along $\mathbb{R}$
and the space coordinate $\sigma$ labels the points on ${\sS}$
which is assumed to have the topology of $S^1$. Therefore, using
\begin{eqnarray}
&& q^{IJ} =q^{IJ}_a d y^a = q^{IJ}_0 d \tau + q^{IJ}_1 d \sigma,
\nonumber\\
&& e^I = e^I_0 d \tau + e^I_1 d \sigma,
\end{eqnarray}
The action becomes
\begin{eqnarray}
&& S = \int_{\mathbb{R}} d \tau \int_{{\sS}} d \sigma \left (
{\dot q}^{IJ}_1 p_{IJ} + {\dot e}^I_1 \pi_I - \lambda^{IJ} {\cal
D}_{IJ} - \lambda^I {\cal G}_I \right ),
\end{eqnarray}
where $\lambda^{IJ}:=- q^{IJ}_0$ and $\lambda^I:= - e^I_0$ are
Lagrange multipliers imposing the constraints
\begin{eqnarray}
&& {D}_{IJ}=  \partial_{\sigma} p_{IJ} \approx 0 \, \\
&& {C}_I = \partial_{\sigma} \pi_I + \varepsilon^{KL}\,_{IJ} e^J_1
p_{KL} \approx 0.
\end{eqnarray}
There are no more constraints. Smearing the constraints with test
fields
\begin{eqnarray}
D(N)= \int_{\sS} d\sigma N^{IJ} {\cal G}_{IJ}, \quad C(a)=
\int_{\sS} d \sigma a^I {\cal C}_I
\end{eqnarray}
to compute the Poisson brackets
\begin{eqnarray}\label{lag}
&&\{ D(N), D(M) \} = 0, \nonumber\\
&&\{ D(N), C(a) \} = 0, \nonumber\\
&&\{ C(a) , C(b) \} = D( \ast [a,b])
\end{eqnarray}
with $[a,b]^{IJ} := a^I b^J - a^J b^I$. Thus all the 10
constraints are first class for the 10 configuration variables
$(q^{IJ}_1, e^I_1)$. Therefore, the system has no local degrees of
freedom, it is a topological field theory.

In the spirit of what was done in the previous section now we
couple this world sheet action to the four-dimensional BF theory
in such a way to maintain the topological character of the model.
There is a natural choice of coupling leading to the new model
introduced in this section, namely:
\begin{eqnarray}\label{bfymgr}
S_{\va BFYMGR}&=& \int_{\sM} B_{IJ}\wedge
F^{IJ}(\omega)+\int_{\sW} \left ( \left [ B^{IJ}{\cal E}_a{\cal
E}^a -d_{\omega} q^{IJ} +\ast (e^I \wedge e^J) \right ] p_{IJ} +
\pi_I d_{\omega} e^I+{\cal E}_aF^a(A)\right ).
\end{eqnarray}
We call this model $BFYMGR$ (where $GR$ stands for general
relativity) due to the suggestive similarity of the equations of
motion with those of general relativity in the first order
formalism. In order to make this statement more explicit let us
analyze the equations of motion of the model. The observation is
that on the world sheet variations with respect to $p$ imply that
$B={\cal E}^{-2} (*(e\wedge e)-dq)$, hence the $B$ field is simple
up to a gauge transformation. Therefore, the simplicity
constraints that reduce BF theory to general relativity are
satisfied on the world sheet. The conclusion is more transparent
is we study the remaining equations of motion. For instance we
have \be F^{IJ}=-p^{IJ} \ {\cal E}^2 \delta_{\sW}\ \ \ \rightarrow
\ \ \ \bar F^{IJ}_{\mu\nu}=-p^{IJ} \ {\cal E}^2 \ \ \ \  {\rm and
} \ \ \ \ \epsilon_{IJKL} e^J p^{KL}=d_A\pi_{I}, \ee where $\bar
F^{IJ}_{\mu\nu}$ is the smearing of the curvature tensor a
two-dimensional surface dual to the world sheet along the
coordinates $\mu-\nu$, more precisely
\[\bar F^{IJ}_{\mu\nu}:=\int_{\mu-\nu} F^{IJ}. \]
Now we can appropriately combine  the previous equations  and
obtain \be \label{einst}\epsilon^{\mu\nu\rho\tau} \epsilon_{IJKL}
e^J_{\nu} \bar F_{\rho\tau}^{KL} = \epsilon^{\mu\nu}(d_A
\pi_{I})_{\nu} {\cal E}^2, \ee where
$\epsilon^{\mu\nu}:=\epsilon^{\mu\nu\rho\tau} (dt)_{\rho}
(d\sigma)_{\tau}$, and we have assumed that ${\cal E}^2$ is non
vanishing in order to bring it to the right hand side. The
previous equation has a suggestive similarity to Einstein's
equation with source $T_{\mu\nu}=t_{I (\mu}e_{\nu)}^I$ where $t_{I
\mu}=(d_A \pi_{I})_{\mu} {\cal E}^2$. This is why we call this
topological model $BFYMGR$.

We have emphasized the similarity of this model with Einstein's
theory of gravity in order to motivate the introduction of this
model. Now let us stress why this is quite different in fact. The
main reason is that, in contrast with general relativity, this
model is a topological theory with no local excitations. This
conclusion becomes transparent in the Hamiltonian analysis which
yields the following set of constraints for the canonical
variables $(E^{\mu}_{IJ}, A_{\nu}^{KL})$, $(p^{IJ}, q_1^{KL})$,
$({\cal E}_a,{\cal A}_1^b)$, and $(\pi_I,e_1^J)$ \ba
\label{nuevos} &&\nonumber L_{IJ}:=d_{A{\mu}}E^{\mu}_{IJ} + 2
\delta_{{\sS}}[ q_{[I|M|} p^M_{\ J]}]\approx 0, \\ &&\nonumber
K^{\mu IJ}:=\epsilon^{\mu\nu\rho}F^{IJ}_{\nu\rho}(x) +
\delta_{{\sS}}[{\cal E}_a{\cal E}^a p^{IJ}\partial^{\mu}_{\sigma}] \approx 0, \\
&&\nonumber {G}_a := d_{\cal A}{\cal E}_a\approx 0 \ea which are
just the same as (\ref{gauss1}), (\ref{curvature}) and
(\ref{cont}) in addition to the new world sheet constraints \ba
\label{einstein} && {C}_I := d_{\omega} \pi_I + 2 e^J\, \ast
p_{IJ}\approx 0 \ea It is easy to see using the results of the
previous sections that the constraints form a first class set of
$24+{\rm dim}(\frak g)$ local constraints for the same number of
configuration variables. The degrees of freedom are topological.

\section{A two-dimensional background independent Yang-Mills
theory}\label{2dym} Using what we have learnt we can also define a
$2$-dimensional background independent Yang Mills theory by making
the 2-form $\rho$ appearing in eq. (\ref{ym}) dynamical in an
world sheet intrinsic way: namely $\rho=(e^I \wedge e^J )p_{IJ}$.
The resulting action is
\begin{eqnarray}\label{tym}
S_{\va TYM}&=& \int_{\sW} \left ( \left [ d q^{IJ} + \ast (e^I
\wedge e^J)+e^I \wedge e^J {\cal {\cal E}}_a{\cal {\cal
E}}^a\right ] p_{IJ} +{\cal E}_aF^a(A)+ \pi_I d e^I \right ).
\end{eqnarray}
The canonical analysis performed along the lines of the one
corresponding to the previous model leads to the following
constraints
\begin{eqnarray}
&& {G}_a = d_A{\cal E}_a\approx 0 \, \\
&& {D}_{IJ} =  \partial_{\sigma} p_{IJ} \approx 0 \, \\
&& {C}_I = \partial_{\sigma} \pi_I + 2 e^J\,_1 p_{IJ} {\cal {\cal
E}}_a {\cal {\cal E}}^a
    + 2 e^J\,_1 \ast p_{IJ}\approx 0
\end{eqnarray}
The first one is the familiar Gauss law of Yang Mills theory while
the remaining ones correspond to the appropriate modification of
the ones obtained above. The constraint algebra gives
\begin{eqnarray}
&&\{ G(\alpha), G(\beta) \} = G([\alpha,\beta]_{\frak{g}}), \nonumber\\
&&\{ D(N), D(M) \} = 0, \nonumber\\
&&\{ D(N), C(a) \} = 0, \nonumber\\
&&\{ C(a) , C(b) \} = D( \ast [a,b]+{\cal E}^2
[a,b])+G(2[a,b]^{IJ} p_{IJ} \ {\cal E})
\end{eqnarray}
with $[a,b]^{IJ} := a^I b^J - a^J b^I$ and
$[\alpha,\beta]_{\frak{g}}$ is the commutator in the Lie algebra
$\frak{g}$. The constraint algebra closes and gives a first class
system. As before we have $10+{\rm dim}(\frak{g})$ local
constraints for the same number of configuration variables; hence
the system is a topological field theory.

We end this section with a remark. Notice that the constraint
algebra has field dependent structure constants. This is
characteristic of the constraint algebra of general relativity,
although here the field dependence is much simpler since the
quantity ${\cal E}^2$ is constant on the world sheet due to the
Gauss constraint. These are genuine field dependent structure
constants.

\section{Quantization}\label{discussion}

We have shown how the coupling of four-dimensional BF theory to strings
introduced in \cite{bp} allows for the definition of a large class of
topological field theories with physically interesting kinematical degrees of
freedom. The set possibilities is indeed very large so we have concentrated
here on two cases of special interest: world sheet Yang-Mills theories defined
in terms of structure groups $G$ possessing an ${\rm ad}_G$ invariant metric
in their Lie algebra $\frak{g}$, and a world sheet tetrad (with intriguing
resemblance with general relativity).

The fact that these models are topological indicates that their
non-perturbative quantization should be well defined. Indeed the
quantization of the model of Section \ref{ymbf} follows
straightforwardly from the results of \cite{bp} and
\cite{winston}. This should be clear from the fact that the phase
space structure presented in Section \ref{ymbf} is quite similar
to the one of the theory briefly reviewed in Section \ref{strings}
whose loop quantization is set up in \cite{bp} and completely
worked out in \cite{winston}. The only new ingredient are the
Yang-Mills unconstrained degrees of freedom which are specially
well suited for the application of loop variables techniques.

\begin{figure}[h]
\centerline{\hspace{0.5cm} \(
\begin{array}{c}
\includegraphics[height=6cm]{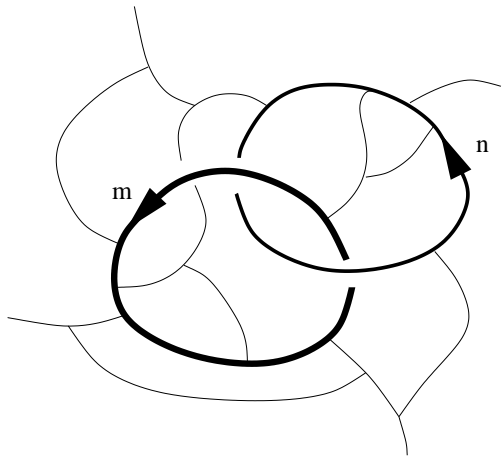}
\end{array}\)} \caption{The elements of a natural basis of the kinematical Hilbert
 can be written as the product of: 1) A functional of the Lorentz connection labelled by a graph in
 space and the assignment of unitary
irreducible representations of the Lorentz group, i.e., a {\em
$SO(3,1)$ spin-network state}
 (represented by the thin-lines graph), 2) An $n$-point
 spin function (represented here by the endpoints of the thin-lines-graph on
 the strings; see \cite{bp} for the precise definition),
3) A functional of the Yang-Mills connection given by the product
of Wilson loops on a unitary representation of the structure group
$G$ along the each string component.} \label{stringy}
\end{figure}

More precisely a basis of the kinematical Hilbert space---space of
solutions of all quantum constraints with the sole exception of
the curvature constraint (\ref{curvature1})---of the model
(\ref{bfym}) is given by: (1) A bulk $S0(3,1)$ spin network
functional of the $S0(3,1)$-connection $A$ based on a graph
$\gamma\in \Sigma$ with open ends at $n$ points on the string
$\sS$, (2) an $n$-point spin functional of $\lambda$ (recall that
the variable $p=\lambda v\lambda^{-1}$ for $v\frak{g}$ normalized
and $\lambda\in G$), (3) a functional of the $G$-connection $\cal
A$ given by the trace of the Wilson loop of $\cal A$ around the
string $\sS$ in an unitary irreducible representation of $G$
(Figure \ref{stringy}). If $G$ is compact we can always think of
the latter quantum number as $n\in \N$, where $n$ labels the
$n$-th eigenvalue $\epsilon_n$ of the square of the electric field
$\widehat{{\cal E}^a{\cal E}_a}$. The physical Hilbert space is
obtained by imposing the quantum version of the constraint
(\ref{curvature1}). This amounts for requiring the holonomy of
loops around the string carrying Yang-Mills quantum flux number
$n\in \N$ to be in the conjugacy class of $\exp{(-\epsilon_n v)}$.
The techniques developed in \cite{winston} can be simply extended
to treat this case.

Another important remark concerns the relationship of this model
with 4-dimensional Yang-Mills coupled to general relativity. There
is a close relationship between $SO(3,1)$ BF theory and general
relativity \cite{pleb,mike,depietri,capo}. More precisely one can
obtain the action of general relativity in the first order
formulation by constraining the $B$ field to be of the form
$B=(e\wedge e)*$ for a tetrad field $e$. This idea is in fact at
the core of the definition of many spin foam models for
four-dimensional quantum gravity \cite{foamy}. Here we would like
to point out that if such constraint is imposed on the $B$ field
appearing in the action (\ref{bfym}) then the naive quantum
amplitude for a world sheet configuration with quantized
Yang-Mills electric field squared $\epsilon_n$ is proportional to
$\exp{( i \ A_p[\sW] \epsilon_n} )$ where $A_p[\sW]$ is the area
of the world sheet computed with the area form $(e\wedge
e)^{*IJ}p_{IJ}$. This is precisely the functional dependence of
the Yang-Mills amplitude in any dimension \cite{Conrady:2006ek}.
We think that the model presented here might present a new
perspective for the definition of a natural coupling of Yang-Mills
fields with gravity in the context of spin foam models of quantum
gravity.

It would be interesting to undertake the quantization of the model
of Section \ref{gr}. This would require the non-perturbative
quantization of the tetrad field $e^I_1$ and its conjugate
momentum $\pi_I$. We would like to study this question in detail
in the future. Nevertheless, it seems clear that topological
invariance should considerable simplify matters. Its seems that if
this question can be resolved then one should be able to quantize
the model of Section \ref{2dym}. An interesting feature of these
models (from loop quantum gravity perspective) is  that their
constraint algebras represent simpler models of that of general
relativity, since as in the latter, they possess field dependent
structure constants. Perhaps some technical issues concerning the
quantization of such theories can be clarified in this simpler
context. The model of Section \ref{gr} is in addition interesting
because of its additional resemblance to general relativity.
\section{Some speculative remarks}\label{especulemos}

Let us finish with more speculative considerations which are
however an important additional motivation for the study presented
here. The most fundamental question of loop quantum gravity is
whether one can construct a quantum field theory in the absence of
a non-dynamical background metric. Several known results such as
the quantization of Chern-Simons theory, 2+1 gravity, BF theory,
etc., show that this is possible at least when dealing with
topological field theories. The difficult question is whether one
can construct an explicit non trivial example of background
independent quantum field theory (with infinitely many degrees of
freedom, i.e., infinitely many physical observables). One can
argue that the entire framework of standard quantum field theory
is based on the notion of {\em particle}, where a Fourier modes
are the basic building block in the construction of standard
quantum field theories. Similarly, we would like to explore the
possibility that the finitely many degrees of freedom encoded in
topological models, of the kind presented here, might be put
together (be `second quantized') in order to define a QFT with
infinitely many degrees of freedom. Our ideas are at this stage
rather heuristic with some aspects based in unproven assumptions
motivated by properties of very simple models \cite{2dg}. The
degree to which these assumptions can be made into factual
statements will be explored elsewhere.

The basic idea goes as follows: In the model of \cite{bp} as well
as those presented here, the topology of the space time manifold
$\sM$ and the embedded world sheet $\sW$ are held fixed. Under
these conditions the transition amplitudes between kinematical
states can be computed. When the topology of the world sheet is
trivial (e.g. a cylinder $\sW=S^1 \times \R$ or an ensemble of any
arbitrary number of disconnected cylinders) these amplitudes can
be used to define the so called physical inner product of the
(canonically defined) quantum theory. Let us call $\sH_n$ with
$n\in \N$, the physical Hilbert space so defined for the quantum
theory associated with classical configuration space containing
$n$ disconnected strings. One can construct a theory with
infinitely many degrees of freedom defining the `Fock' space
$\sF=\oplus_{n=0}^{\infty}\sH_n$ with the infinite set of quantum
observables associated with the multi-string states (for the
explicit construction in the particle case see
\cite{Noui:2006kv}). However, from our perspective \footnote{In
\cite{Noui:2006kv} the context in which $\sF$ is introduce is
quite different. There one uses it to setup a perturbation
theory.} such a theory seems rather trivial because there is no
interaction between the $\sH_n$'s for different values of $n$.
\begin{figure}[h]
\centerline{\hspace{0.5cm} \(
\begin{array}{c}
\includegraphics[height=3cm]{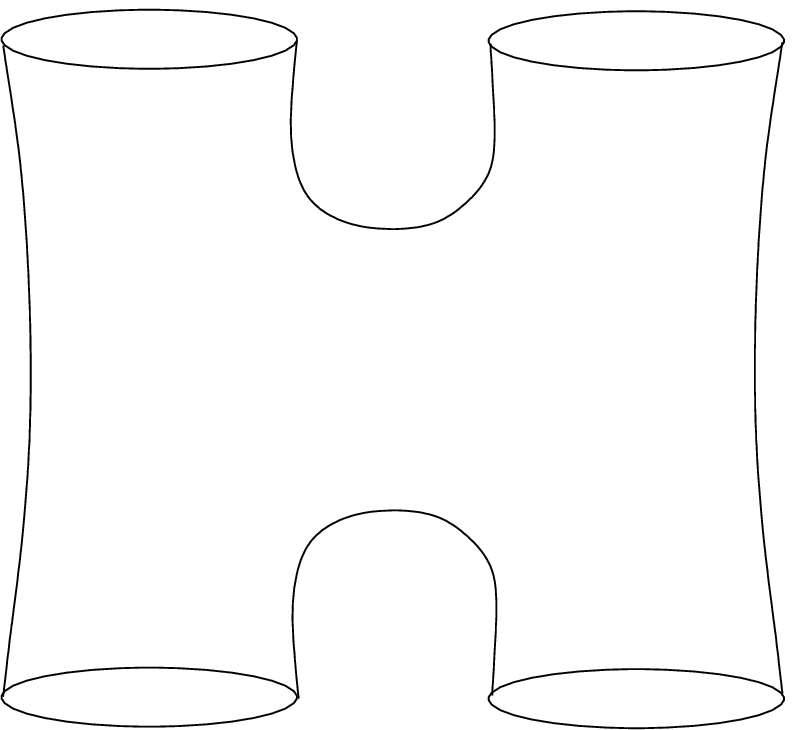}
\end{array}\ \ \ \ \ \ \ \ \ \ \ \ \ \ \ \ \
\begin{array}{c}
\includegraphics[height=3cm]{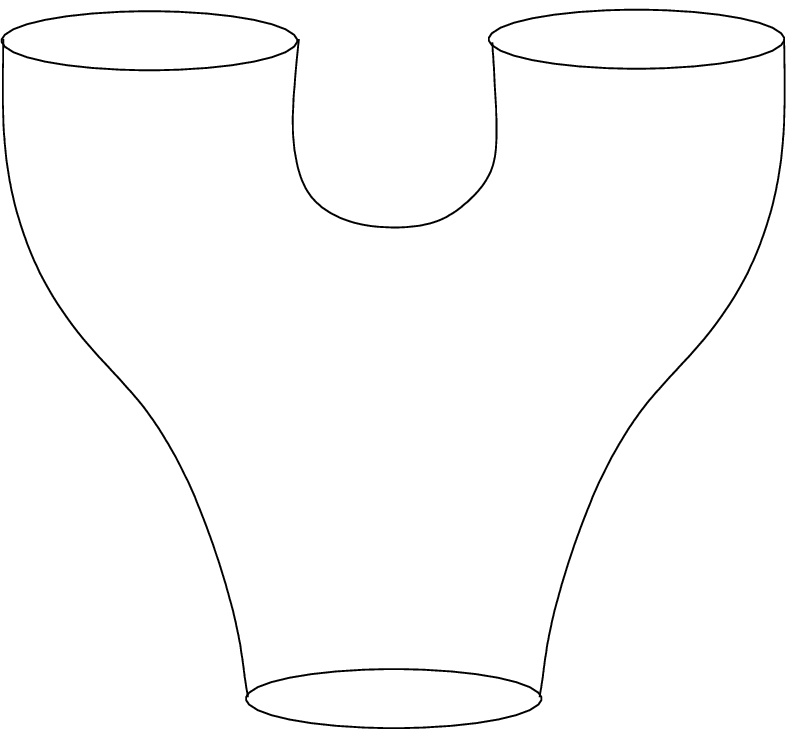}
\end{array}\)}
\caption{Interacting string world sheets} \label{ws}
\end{figure}

When the world sheet topology is non trivial (e.g. it has
branching components  as in Figure \ref{ws} and/or non vanishing
genus) the quantum amplitudes are still well defined (in the spin
foam representation) but have no clear-cut physical interpretation
\footnote{A field theoretic interpretation as Feynman diagrams in
the context of perturbation theory of an associated effective
field theory is proposed in \cite{lau}.}. It is tempting to
interpret these amplitudes as providing the definition of physical
interacting transition amplitudes in a theory where the
kinematical Hilbert space is the Hilbert space $\sF$ defined
above. This interpretation would be consistent if: (1) the sum
over world sheet topologies would be convergent, and (2) the
transition amplitudes define a positive semidefinite inner product
in $\sF$. This last requirement is highly non trivial---it is the
counterpart of unitarity in background dependent quantum field
theory. If these conditions hold, this would provide a consistent
way of rendering the world sheet topology dynamical achieving the
goal of defining a non trivial (i.e. interacting) quantum field
theory with infinitely many degrees of freedom: the latter given
by the ensemble global degrees of freedom of all world sheet
topologies.

Due to the fact that topology of two-dimensional orientable
manifolds is characterized by a single integer (the genus $g$)
condition (1) above can be satisfied if the amplitudes are
suitably damped for high $g$. In fact the sum over two-dimensional
topologies does converge in simple models such as 2d BF theory
(see for instance \cite{2dg}). Some positive indication that
property (2) could be realized for models of the kind presented
here also comes from the study of this simple case. However, the
model in \cite{2dg} is too simple an the sum over world sheets
does not lead to a theory with infinitely many degrees of freedom.
If the sum over world sheet topologies can be achieved in the
models presented here, due to the the non trivial character of the
degrees of freedom involved, be believe they might lead to non
trivial examples of background independent field theories with
infinitely many degrees of freedom. We would like to explore this
possibility in the future.

\section{Discussion}

We have shown how the extended nature of the conical defects that
naturally couple to four dimensional BF theory allow for the
introduction of physically interesting world-sheet fields  while
keeping the topological character of the theory. These models are
expected to be non-perturbatively quantizable. In particular, the
coupling of Yang-Mills theory with BF theory described in Section
\ref{ymbf} can be quantized in a rather direct way by using the
thecniques of Refs. \cite{bp,winston}. For this theory we get at a
remarkably simple description of states in the kinematical Hilbert
space where bulk-geometry spin network states are dual to
Yang-Mills electric field flux lines (see Figure \ref{stringy}).
The strength of the conical singularities at the location of flux
lines is proportional to the electric field square.

The models are in close relationship with gravity in at least two
independent ways. On the one hand, as we argued in Section
\ref{geom}, solutions of the topological models are in one to one
correspondence with solutions of Einstein's equations. This
correspondence between solutions has to be interpreted with due
care as the gauge symmetries of our models is much larger than the
one of general relativity. In particular local excitations such as
gravitons are pure gauge in our model. Nevertheless the
correspondence among solutions might be of relevance if some of
the hopes described in the previous section could be realized. On
the other hand, our model is linked to gravity along the well
known relationship between four dimensional BF theory and general
relativity explicitly exhibit in the Plebansky formulation of
gravity. In particular, it would be interesting to compare our
model with the coupling to Yang-Mills theories proposed in
\cite{Oriti:2002Bn}.

These models are simple but non trivial. In particular, the
presence of geometric degrees of freedom as well as matter-like
degrees of freedom make them potentially useful for the study of
various conceptual difficulties in non-perturbative quantum
gravity.

\section*{Acknowledgements}

This work was supported in part by: the Agence Nationale de la Recherche, Grant No. ANR-06-BLAN-0050, the 
Coordena\c{c}\~{a}o de Aperfei\c{c}oamento de Pessoal de 
N\'{\i}vel Superior (Capes) through a visiting professor fellowship,
and by CONACyT, M\'exico, Grant No. 56159-F.

\section{Appendix}

Here we construct a truly background independent model which will
lead to a genuinely topological theory. The discussion of the
first part of this paper gives a clear way to defining a
background independent analog. The action is
\begin{eqnarray}\label{toy2}
S&=& \int_{\sW} \left [ {\cal E}_{a} F^{a}(A)+ (\beta \ (e^I
\wedge e^J)^*+\gamma\  e^I \wedge e^J) p_{IJ }{\cal E}_{a}{\cal
E}^{a} +p_{IJ}F^{IJ}(\omega) + \pi_I d_{\omega} e^I \right ],
\end{eqnarray} where we have replaced $q$ by $q=\ast (e^I \wedge
e^J)p_{IJ }$ in the previous action, $F= \left ( \frac12
F^a\,_{\mu\nu} d x^{\mu} \wedge d x^{\nu} \right ) \otimes J_a$
with
$F^{a}_{\mu\nu}(A)=\partial_{\mu}A^a_{\nu}-\partial_{\nu}A^a_{\mu}+f^{a}_{\
bc}A^b_{\mu}A^c_{\nu}$ and---in order to make the $e^I$ and
$p_{IJ}$ fields dynamical---added the natural term
$p_{IJ}F^{IJ}(\omega) +\pi_I d_{\omega} e^I$ which also requires
the introduction of the connection $\omega^{IJ}$. Of course there
are other additional fields which can be added to the action
(\ref{toy2}) but, for the moment, let us look just at this action.
The parameters $\beta$, and $\gamma$ are coupling constants.

After the 1+1 decomposition, $(x^{\mu})=(x^1,x^2)=(\tau,\sigma)$,
each of the terms become: The action becomes (neglecting space
boundary terms)
\begin{eqnarray}
S= \int d \tau \wedge d \sigma \left[  {\cal E}_a {\dot A}^a\,_1 +
\pi_{IJ}{\dot \omega}^{IJ}\,_1 + p_I {\dot e}^I\,_1 - \lambda^a
G_a - \lambda^I C_I - \lambda^{IJ} D_{IJ} \right ]
\end{eqnarray}
with
\begin{eqnarray}
G_a &:=& d_A {\cal E}_a \\ \nonumber C_I &:=& d_{\omega} \pi_I +
\beta p_{IJ}^* e^J\,_1 {\cal E}_a {\cal E}^a +
\gamma p_{IJ} e^J\,_1 {\cal E}_a {\cal E}^a\\
\nonumber D_{IJ} &:=& d_{\omega} p_{IJ} + \frac12 \left ( \pi_I
e_{J1}- \pi_J e_{I1} \right ).
\end{eqnarray}
Smearing the constraints with test fields
\begin{eqnarray}
G(\alpha):= \int_{{\sS}} d \sigma \alpha^a G_a, \quad C(\lambda):=
\int_{{\sS}} d \sigma \lambda^I C_I, \quad D(N):= \int_{{\sS}}d
\sigma N^{IJ} D_{IJ}
\end{eqnarray}
The constraint algebra gives

\begin{eqnarray}
\{ C(\lambda), C(\Lambda) \}= \int_{\sS} \left ( [\lambda,
\Lambda]^{IJ} \left ( \frac{2 \gamma}{\phi^2} {\cal E}_a {\cal
E}^a \right ) D_{IJ} + \frac{4 \gamma}{\phi^2} {\cal E}^a
[\lambda, \Lambda]^{IJ}  p_{IJ} G_a \right ), \quad
[\lambda,\Lambda]^{IJ}:= \frac12 \left ( \lambda^I \Lambda^J -
\lambda^J \Lambda^I \right )
\end{eqnarray}
\begin{eqnarray}
\{G(\alpha), G(\beta)\}&=&G([\alpha,\beta]) \\ \nonumber
\{G(\alpha), C(M)\}&=&\{G(\alpha), D(N)\}=0\\ \nonumber \{D(N),
D(M)\}&=&D([N,M])\\ \nonumber \{D(N), C(\beta)\} &=& C(N\cdot
\beta)\\ \nonumber \{ C(M), C(N) \}&=&{\cal E}^2 D( [N,M])+2
G([N,M]\cdot p \ {\cal E})
\end{eqnarray}
where $[N,M]^I$ and $(N\cdot \beta)^I:=N^{IJ}\beta_J$ is the
commutator in the Lie algebra $so(4)$. The constraints are all
first class which leads to the conclusion that there are zero
local degrees of freedom.

\end{document}